\documentclass[journal]{IEEEtran} 
\usepackage[acronym,shortcuts,nonumberlist,hyperfirst=false,nopostdot,nogroupskip,nomain]{glossaries}
\usepackage{cite}
\usepackage[pdftex]{graphicx}
\graphicspath{{./figures}}
\DeclareGraphicsExtensions{.pdf,.jpeg,.png} 
\usepackage{tikz,tikz-3dplot}
\usetikzlibrary{arrows, automata, shapes.geometric, positioning, decorations.pathreplacing, calc, matrix}
\usepackage{amsmath,bm}
\usepackage{amssymb}
\usepackage{comment}
\usepackage{algorithmic}
\usepackage{algorithm}
\usepackage{caption}
\usepackage{subcaption}
\usepackage{array}
\usepackage{mathtools}

\DeclareMathSymbol{\shortminus}{\mathbin}{AMSa}{"39}
\usepackage{readarray}
\usepackage[bookmarks=false]{hyperref}

\newcommand{\Ktot}{K_{\mathrm{tot}}}

\newcommand{\M}[4]{\Delta_{#1\rightarrow#2}^{#3}(#4)}

\newcommand{\CN}[3]{\mathcal{CN}(#1;#2,#3)}
\newcommand{\OC}[1]{\mathcal{O}(1)}

\newcommand{\Tr}{\mathrm{T}}

\newignoredglossary{ignore}

\newacronym{BiGAMP}{BiGAMP}{bilinear generalized approximate message passing}
\newacronym{SISO}{SISO}{soft-input-soft-output}
\newacronym{SPARC}{SPARC}{sparse regression code}
\newacronym{BiSPARC}{BiSPARC}{bilinear sparse regression code}
\newacronym{URA}{URA}{unsourced random access}
\newacronym{RV}{RV}{random variable}
\newacronym{SIC}{SIC}{successive interference cancellation}

\newacronym{CLT}{CLT}{central-limit theorem}

\newacronym{IoT}{IoT}{Internet-of-Things}
\newacronym{mMTC}{mMTC}{massive machine type communications}
\newacronym{HTC}{HTC}{human-type communications}
\newacronym{MAC}{MAC}{Multiple Access Channel}
\newacronym{MMSE}{MMSE}{Minimum Mean Squared Error}
\newacronym{LMMSE}{LMMSE}{Linear Minimum Mean Squared Error}
\newacronym{AMP}{AMP}{Approximate Message Passing}
\newacronym{BP}{BP}{Belief Propagation}
\newacronym{pmf}{pmf}{probability mass function}
\newacronym{LDPC}{LDPC}{low-density parity check}
\newacronym{mMIMO}{mMIMO}{massive MIMO}
\newacronym{MIMO}{MIMO}{multiple-input-multiple-output}
\newacronym{CCS}{CCS}{coded compressed-sensig}
\newacronym{MRC}{MRC}{maximum ratio combining}

\newacronym{SCL}{SCL}{successive cancellation list}
\newacronym{CRC}{CRC}{cyclic redundancy check}
\newacronym{SCMA}{SCMA}{sparse code multiple access}
\newacronym{NOMA}{NOMA}{non-orthogonal multiple access}

\def\-{\raisebox{.75pt}{-}}
\begin{document}
%
\title{BiSPARCs for Unsourced Random Access in Massive MIMO}
\author{
  \IEEEauthorblockN{
    Patrick Agostini$^{*\dagger}$, Zoran Utkovski$^{*}$, S\l{}awomir Sta\'{n}czak$^{*\dagger}$
  }\\
  \IEEEauthorblockA{
    Fraunhofer Heinrich-Hertz-Institute Berlin, Germany$^{*}$\\    
    Technische Universität Berlin, Germany$^{\dagger}$\\
    \{patrick.agostini, zoran.utkovski, slawomir.stanczak\}@hhi.fraunhofer.de
  }
}
%
\maketitle
%

\begin{abstract}
    This paper considers the massive MIMO unsourced random access problem in a quasi-static Rayleigh fading setting. The proposed coding scheme is based on a concatenation of a "conventional" channel code (such as, e.g., LDPC) serving as an outer code, and a sparse regression code (SPARC) serving as an inner code. The scheme combines channel estimation, single-user decoding, and successive interference cancellation in a novel way. The receiver performs joint channel estimation and SPARC decoding via an instance of a \ac{BiGAMP} based algorithm, which leverages the intrinsic bilinear structure that arises in the considered communication regime. The detection step is followed by a per-user soft-input-soft-output (SISO) decoding of the outer channel code in combination with a \ac{SIC} step. We show via numerical simulation that the resulting scheme achieves stat-of-the-art performance in the massive connectivity setting, while attaining comparatively low implementation complexity.
\end{abstract}


\section{Introduction}

\IEEEPARstart{M}assive machine type communication is envisioned to constitute an important aspect of the technological evolution towards future 6G networks. Different from human-centric communication, \ac{mMTC} 
services are characterized by the presence of a potentially massive number of users that transmit short packets in a sporadic fashion and potential applications are in various domains, ranging from industry and smart cities to logistics and healthcare. The traditional “grant-based” random access strategy employed in current network solutions, requires a terminal identification and resource granting step prior data transmission. The signaling involved in the granting process requires a unique combination of pilot/signature signal and time-frequency slot for each device, such that the base station is able to identify the users requesting access. Due to the entailed protocol overhead, and the necessity for individual and distinct signatures, this approach does not scale well in the \ac{mMTC} regime. The problem is further exacerbated by the reduced efficiency of channel codes in the finite block-length (FBL) regime. In this context, it is desirable to let users transmit without any prior resource request, i.e., the active users have to be simultaneously detected and decoded within a single communication phase. Such grant-free access requires a departure from the design assumptions prevailing in current cellular systems, which has sparked the development of novel random access schemes \cite{wu_massive_2020}. 


From the plethora of various grant-free random access protocols, \acf{URA} \cite{polyanskiy_perspective_2017} has recently attracted considerable attention. According to this paradigm, users employ the same codebook to transmit information thereby separating the problem of user identification from the actual data transmission. The decoder only declares which messages were transmitted, without associating the messages to the users that transmitted them, shifting this task to the higher layers. In the \ac{URA} paradigm, the explicit initial access phase is bypassed, which decouples the system complexity from the total number of users, allowing for full system scalability. These features make it well suited to the envisioned \ac{mMTC} scenarios.

%
%
%

\subsection{Related Work}

Since its initial conceptualization in \cite{polyanskiy_perspective_2017}, the \ac{URA} paradigm has attracted considerable interest from the research community, giving rise to many candidate solutions tailored to various communication scenarios. Earlier works such as \cite{polyanskiy_perspective_2017} focus on  information-theoretic limits and propose coding schemes for the additive white Gaussian noise (AWGN) channel setting. Later on, several practical schemes based on compressed sensing have been proposed for the same setting   \cite{fengler_sparcs_2020},  \cite{pradhan_joint_2019}. An extension to the quasi-static Rayleigh fading channel  has been addressed in \cite{kowshik_energy_2020}, where a scheme based on \ac{LDPC} codes using a belief propagation decoder has been proposed. 

More recently the focus has shifted towards coding schemes for the \ac{mMIMO} regime. 
Initially, the \ac{URA} problem on a Rayleigh block-fading channel in a \ac{mMIMO} setting was formulated in \cite{fengler_non-bayesian_2021}. The authors combined the \ac{CCS} paradigm with an outer tree code \cite{amalladinne_coded_2018}, and a covariance-based activity detection (AD) algorithm, showing the capability of achieving sum spectral efficiencies that scale with the coherence blocklength.  

In the operating regime where the coherence blocklength exceeds the number of active users, the state-of-the-art approaches for \ac{URA} leverage the large blocklength to device two-step approaches, in which communication resources are split between   user activity detection and channel estimation on one hand, and coherent detection on the other hand. Specifically,  \cite{fengler_pilot-based_2022} proposed the first pilot-aided communication scheme that employs multi-measurement-vector approximate message passing (MMV-AMP) for user activity detection and channel estimation, in combination with \ac{MRC} outer Polar channel coding. Similar in spirit, a pilot-based scheme with symbol spreading has been proposed in \cite{gkagkos_fasura_2022} capable of out-performing \cite{fengler_pilot-based_2022} in the considered communication regimes, but at the cost of increased computational complexity. In \cite{ahmadi_unsourced_2022}, a multiple stage approach of orthogonal pilots appended with polar code encoded messages with low complexity receiver has been proposed. For completeness, we also  mention the tensor-based modulation (TBM) scheme proposed in \cite{decurninge_tensor-based_2020}, which employs non-coherent tensor modulation with efficient canonical polyadic decomposition (CPD) based detection. Despite being less competitive than the coherent pilot-aided counterparts, this approach is of interest due to the fact that it leverages matrix/tensor factorization, thus sharing  conceptual similarities with our approach presented hereafter.

\subsection{Main Contributions}

In this paper, we propose a scheme for \ac{URA} in the quasi-static fading setting. The proposed scheme combines a \acp{SPARC}-based inner code and an outer channel code, leveraging probabilistic Bayesian detection/decoding to operate in the massive MIMO \ac{URA} regime. 
Similar to \cite{gkagkos_fasura_2022} and \cite{fengler_pilot-based_2022}, the receiver runs an iterative procedure composed of one detection and decoding step, followed by a \ac{SIC} step. The proposed scheme leverages the bilinear relation between the MIMO channels and the binary \ac{SPARC} support-vectors which arises in the considered regime to perform a joint \ac{SPARC} support-vector and channel estimation (hence BiSPARCs). We derive a probabilistic detection procedure using the \ac{BiGAMP} paradigm with message updates that consider the structure of the \acp{SPARC} encoding structure as prior. The probabilistic detection of the support-vectors allows for parallel  soft-decoding of the outer channel code for each of the individual  messages. This parallel decoding yields an implicit user separation, which mitigates the need for extra parity-check based outer-coding, as widely employed in compressed-sensing based \ac{URA} schemes \cite{amalladinne_coded_2018,fengler_sparcs_2020}. Successfully decoded messages are re-encoded for a subsequent \ac{SIC} step in which their contribution together with the estimated MIMO channels are removed from the receive signal. The residual signal is then fed into the detection step for the subsequent round of detection and decoding steps, as illustrated in Fig.~\ref{fig:scheme}. 

The proposed iterative detection scheme shares similarities with the previously introduced schemes \cite{fengler_pilot-based_2022, gkagkos_fasura_2022}, with the major difference being in the "blind" separation between detection (involving joint user activity detection and channel estimation)  and decoding, enabled by the present bilinear structure. In addition, the use of \ac{SPARC} as an inner code allows for non-coherent operation without the need for explicit channel estimation, which frees resources for communication. A further advantage of the proposed scheme is that it does not require additional coding and signaling for user separation as this is provided intrinsically by the joint detection and decoding of the outer channel code. The numerical evaluation of the scheme shows competitive performance with respect to the best performing approaches from the literature at similar or lower computational cost. 

\subsection{Notation}

Unless specified otherwise, we use lower- and upper-case bold letters to denote vectors and matrices, respectively. Upper-case calligraphic letters denote sets. $\mathbf{I}$ denotes the identity matrix of corresponding dimension. We use $\mathbf{x}\sim\mathcal{CN}(\mathbf{x};\boldsymbol{\mu},\boldsymbol{\Sigma})$ to denote that the random vector $\mathbf{x}$ follows a circularly-symmetric complex Gaussian distribution with mean vector $\boldsymbol{\mu}$ and covariance matrix $\boldsymbol{\Sigma}$. We use $\{\cdot\}^{\mathrm{T}}$ and $\{\cdot\}^{\mathrm{H}}$ for the transpose and Hermitian operators, respectively. By $[N]:=\{1,\ldots,N\}$ we denote the counting set of cardinality $N$.
\begin{figure*}[hbt!]
    \centering
    \begin{tikzpicture}[node distance=.8cm]
    \tikzstyle{box} = [
    rectangle, 
    minimum width=1.cm, 
    minimum height=.5cm, 
    text centered, 
    draw=black, 
    rounded corners];
    \tikzstyle{arrow} = [thick,->,>=stealth]
    \node (W1)    [box]                              {\small $f_{\mathrm{out}}(W_1)$};
    \node (W2)    [box, below of=W1, yshift=.0cm]    {\small $f_{\mathrm{out}}(W_2)$};
    \node (dots2) [below of=W2, yshift=.35cm]        {\small $\vdots$};
    \node (WK)    [box, below of=dots2, yshift=.2cm] {\small $f_{\mathrm{out}}(W_K)$};
    \node (EE1)   [box, right of=W1, xshift=1.2cm] {\small $f_{\mathrm{in}}(\mathbf{c}_1)$};
    \node (EE2)   [box, right of=W2, xshift=1.2cm] {\small $f_{\mathrm{in}}(\mathbf{c}_2)$};
    \node (dots4) [below of=EE2, yshift=.35cm]     {\small $\vdots$};
    \node (EEK)   [box, right of=WK, xshift=1.2cm] {\small $f_{\mathrm{in}}(\mathbf{c}_K)$};
    \draw[stealth-, thick] (W1.west) -- ++(-.55,0)node[anchor=center, xshift=-.25cm] {\small $W_1$};
    \draw[stealth-, thick] (W2.west) -- ++(-.55,0)node[anchor=center, xshift=-.25cm] {\small $W_2$};
    \draw[stealth-, thick] (WK.west) -- ++(-.55,0)node[anchor=center, xshift=-.25cm] {\small $W_K$};
    \draw[arrow] (W1.east) -- (EE1.west);
    \draw[arrow] (W2.east) -- (EE2.west);
    \draw[arrow] (WK.east) -- (EEK.west);
    \node [box, right of=EE2, xshift=1.5cm] (MAC) {
        \small
        \begin{tabular}{c} 
            mMIMO \\ AP 
        \end{tabular}
    };
    \node (JDEC) [rectangle, xshift=1.5cm, right of=MAC, text centered, draw=black, rounded corners]{\small$g_{\mathrm{in}}(\mathbf{Y})$}; 
    \node (g1) [box, right of=W1, xshift=8cm]       {\small $g_{\mathrm{out}}(\mathbf{c}_1)$};
    \node (g2) [box, below of=g1]                   {\small $g_{\mathrm{out}}(\mathbf{c}_2)$};
    \node (dots5) [  below of=g2, yshift=.35cm]     {\small $\vdots$};
    \node (gK) [box, below of=dots5, yshift=.2cm]   {\small $g_{\mathrm{out}}(\mathbf{c}_K)$};
    \draw[arrow] (MAC.east) -- node[anchor=center, above] {\small$\mathbf{Y}$}(JDEC.west);
    \draw[-stealth,thick] (g1.east) -- ++(.5,0) node[anchor=center, xshift=.25cm] {\small$\hat{W}_1$};
    \draw[-stealth,thick] (g2.east) -- ++(.5,0) node[anchor=center, xshift=.25cm] {\small$\hat{W}_2$};
    \draw[-stealth,thick] (gK.east) -- ++(.5,0) node[anchor=center, xshift=.25cm] {\small$\hat{W}_K$};
    \draw[-stealth,thick] (JDEC.east) -- ++(.25,0) node[anchor=center, above, xshift=.25cm,yshift=.75cm] {\small$\hat{\mathbf{c}}_1$} |- (g1.west);
    \draw[-stealth,thick] (JDEC.east) -- ++(.25,0) node[anchor=center, above, xshift=.25cm,yshift=.0cm] {\small$\hat{\mathbf{c}}_2$} |- (g2.west);
    \draw[-stealth,thick] (JDEC.east) -- ++(.25,0) node[anchor=center, above, xshift=.25cm,yshift=-1.1cm] {\small$\hat{\mathbf{c}}_3$} |- (gK.west);
    \draw[arrow] (EE1.east) -- node[anchor=center, above] {\small$\mathbf{s}_1$} +(.8,0) -| (MAC.north);
    \draw[arrow] (EE2.east) -- node[anchor=center, above] {\small$\mathbf{s}_2$} +(.5,0) -- (MAC.west);
    \draw[arrow] (EEK.east) -- node[anchor=center, above] {\small$\mathbf{s}_K$} +(.8,0) -| (MAC.south);
\end{tikzpicture}
    \caption{This block diagram offers a synopsis of the overall comunication and coding scheme.\label{fig:encoding_scheme}}
\end{figure*}
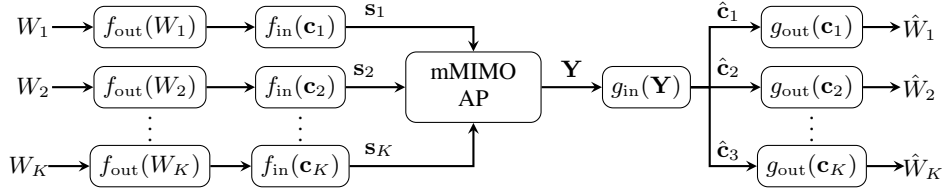
\section{System Model} \label{sec:system_model}

Consider the uplink of a wireless network with $\Ktot$ single-antenna devices and a single receiver equipped with $M$ antennas. We target a communication scenario in which active devices experience quasi-static Rayleigh fading channels over blocks of $T$ signal dimensions in which the user channel vectors are constant. In every transmission block, only a random set $\mathcal{K}$ consisting of $K\ll \Ktot$ devices are simultaneously active. We assume that each active device transmits a message of $B$-bits, $W_{k}\in[2^B], \forall k\in[K]$, where, for convenience, we have labeled active users using integers from the set $\mathcal{K}=[K]$. We assume that the \textit{activity} status of devices is unknown at the receiver. 

Following the unsourced random access paradigm from \cite{polyanskiy_perspective_2017}, we assume that all devices are equipped with a common set of $N$ sequences of length $T$, i.e. a shared dictionary $\mathbf{A}=[\mathbf{a}_1,\ldots,\mathbf{a}_N]\subset \mathbb{C}^{T\times N}$. To transmit data, active devices employ a common encoding function $f:[2^B]\rightarrow\{0,1\}^N$ which maps the information messages $W_k$ to a subset of selected sequences (i.e. dictionary elements) from $\mathbf{A}$, which is realized by the binary selection vector $\mathbf{c}_k\in\{0,1\}^N$, i.e.,
\begin{equation}
    f(W_k)=\mathbf{c}_k.
\end{equation}
We assume that the receiver is equipped with a matching decoding function $g:\{0,1\}^N\rightarrow [2^B]$, i.e., $g(\mathbf{c}_k)=W_k$. The transmit codeword $\mathbf{s}_k\in\mathbb{C}^T$ is generated by linearly combining the selected sequences, i.e.,
\begin{equation}
	\mathbf{s}_k=\sqrt{\alpha_k}\mathbf{A}\mathbf{c}_k,\label{eq:coding1}
\end{equation} 
where $\alpha_k=P_k/\Vert\mathbf{A}\mathbf{c}_k\Vert_2^2$ ensures that the power budget of the $k$-th active device $P_k$ is met. For convenience in the following we will assume that all devices have equal power budget $P_1=\ldots=P_{K}=P$. 

Let $\mathbf{h}_k\in\mathbb{C}^M$ denote the channel vector from the $k$-th active device to the $M$ antennas of the receiver. We assume that $\mathbf{h}_k$ remains constant over a block size of $T$ channel uses. We further assume that $\mathbf{h}_k$ is unknown apriori to both the transmitters and the receiver. Let furthermore $\mathbf{W}\in\mathbb{C}^{T\times M}$ denote the noise realization at the receive antennas where we assume AWGN, i.e., $w_{tm}\sim\mathcal{CN}(0,\sigma^2)$. Assuming that active devices are block synchronous, the baseband representation of the receive signal $\mathbf{Y}\in\mathbb{C}^{T\times M}$ reads
\begin{equation}
\mathbf{Y}=\sqrt{\alpha}\sum_{k=1}^{K}\mathbf{s}_k\mathbf{h}_k^{\mathrm{T}}+\mathbf{W}=\sqrt{\alpha}\mathbf{A}\mathbf{C}\mathbf{H}^{\mathrm{T}}+\mathbf{W},\label{eq:system_model}
\end{equation}
with $\mathbf{H}=[\mathbf{h}_1,\ldots,\mathbf{h}_{K}]\in\mathbb{C}^{M\times K}$ and $\mathbf{C}=[\mathbf{c}_1,\ldots,\mathbf{c}_{K}]\in\{0,1\}^{N\times K}$. Given the received signal $\mathbf{Y}$, the task of the receiver is it to detect the set of binary vectors $\hat{\mathrm{C}}=[\hat{\mathbf{c}}_{1},\ldots,\hat{\mathbf{c}}_{K}]\in\{0,1\}^{N\times K}$ and output a list of valid messages, i.e., 
\begin{equation}
	\hat{\mathcal{L}}:=\{g(\hat{\mathbf{c}}_1),\ldots,g(\hat{\mathbf{c}}_{K})\}=\{\hat{W}_1,\ldots,\hat{W}_{K}\}.
\end{equation}
As a suitable error metric for the decoder, we consider the average per-user probability of error \cite{polyanskiy_perspective_2017} denoted by
\begin{equation}
	P_e=\mathbb{E}\left[\frac{\vert\mathcal{L}\backslash \hat{\mathcal{L}}\vert}{\vert\mathcal{L}\vert}\right],
\end{equation}
where $\mathcal{L}=\{W_1,\ldots,W_{K}\}$ denotes the set of transmitted messages and the expectation is taken over the random choices of codewords, the fading, and the noise. 
The performance of the system is typically measured in terms of the required $E_b/N_0$ for a target $P_e$. 

\section{BiSPARC}\label{sec:encoding}

We will now present the proposed scheme, referred to as \ac{BiSPARC} for \ac{URA}. First, we will describe the encoding and transmission procedure, followed by the detection and decoding steps. The proposed encoding approach comprises an outer coder channel code and an inner coder using \ac{SPARC} coding. The outer channel code is implemented using a suitable off-the-shelf code and concatenated with the inner coder. On the other hand, the decoding procedure involves a message passing-based decoding of the inner \ac{SPARC} code, followed by single-user decoding of the outer channel code. To provide an overview of the encoding and decoding steps, we have included a synopsis of the approach in Fig.~\ref{fig:encoding_scheme}.

\subsection{Encoding}

Consider the general encoding model \eqref{eq:coding1} according to which each active device maps its information message $W_k\in[2^B]$ to the binary support vector $\mathbf{c}_k\in\{0,1\}^N$, which is then mapped to the transmit vector $\mathbf{s}_k$. The linear structure of \eqref{eq:system_model} allows for an interpretation as a concatenation of an inner point-to-point channel $\mathbf{X}\rightarrow \mathbf{A}\mathbf{X}+\mathbf{W}$, with $\mathbf{X}=\mathbf{C}\mathbf{H}^{\mathrm{T}}$, and an outer binary input multiple access channel $(\mathbf{c}_1,\ldots, \mathbf{c}_{K}) \rightarrow \mathbf{X}$. Following \cite{fengler_sparcs_2020}, we will refer to those as the inner and outer channel, the corresponding encoder and decoder will be referred to as inner and outer encoder/decoder. 

\noindent\textit{1) Outer Code:} To code for the outer channel, we rely on a standard channel code, which can be realized by any performant channel code, i.e., \ac{LDPC} or Turbo Codes. In the following, we refer to the outer code via the mapping $f_{\mathrm{out}}:[2^B]\rightarrow [2^{Lm}]$, i.e.,
\begin{equation}
    f_{\mathrm{out}}(W_k)=C_k,
\end{equation}
where $C_k\in [2^{Lm}]$, and $Lm>B$. 
The channel code serves a dual purpose: firstly, the outer code provides error correction against noise and interference, and secondly, it provides an implicit signature unique to each device with high probability, which aids in the user separation process. The outer-code can be realized via any state-of-the-art channel code suitable for the considered code regime. In the remaining, we will mainly present results for 5G new radio (NR) compliant binary \ac{LDPC} code constructions , with sum-product decoding \cite{3gpp_nr_2017}. \smallskip

\noindent\textit{2) Inner Code:} Given the output of the outer encoder $C_k$, the task of the inner code is to modulate the encoded messages into a transmit codeword $\mathbf{s}_k\in\mathbb{C}^T$ which is then transmitted over the shared communication resources. To this end, we resort to a \ac{SPARC} based code construction \cite{fengler_sparcs_2020}, which we denote by $f_{\mathrm{in}}:[2^{Lm}]\rightarrow \mathbb{C}^T$, i.e.,
\begin{equation}
    \mathbf{s}_k=f_{\mathrm{in}}(C_k).\label{eq:mapping1}
\end{equation}
The \ac{SPARC} coder proceeds by first splitting the outer codeword, i.e., $C_k\in[2^{Lm}]$, into $L$ $m$-bit sized chunks denoted by
\begin{equation}
C_k\rightarrow\{C_k^{(1)},\ldots,C_k^{(L)}\},\label{eq:code_chunks}
\end{equation}
where $C_k^{(l)}\in[Q]$, with $Q=2^m$. Given \eqref{eq:code_chunks}, each chunk is encoded into a respective binary vector $\mathbf{c}_k^{(l)}=[c_{1k}^{(l)},\ldots,c_{Qk}^{(l)}]\in\{0,1\}^{Q}$, i.e.,
\begin{equation}
    c_{qk}^{(l)}=\begin{cases}
        1&q=C_k^{(l)}\\
        0&\mathrm{otherwise}.
    \end{cases}\label{eq:subvetor}
\end{equation}
The binary sub-vectors from \eqref{eq:subvetor} are then concatenated to build SPARC-support vector denoted by $\mathbf{c}_k\in\{0,1\}^N$, i.e., 
\begin{equation}
	\mathbf{c}_k:=[\mathbf{c}_k^{(1)};\ldots;\mathbf{c}_k^{(L)}]\label{eq:sparc_coder},
\end{equation}
Finally, the encoder proceeds by mapping the binary \ac{SPARC}-support vector onto the shared dictionary, i.e., $\mathbf{s}_k=\mathbf{A}\mathbf{c}_k$.

\begin{figure*}[hbt!]
    \centering
    \begin{tikzpicture}[node distance=3.5cm]
    \tikzstyle{box} = [rectangle, rounded corners, minimum width=2cm, minimum height=.7cm, text centered, draw=black]
    \tikzstyle{arrow} = [thick,->,>=stealth]
    \tikzstyle{line} = [draw, -latex']

    \node (sic) [box] {\small SIC};
    \node (bigamp) [box, right of=sic] {\small BiGAMP};
    \node (siso1) [box, right of=bigamp, yshift=.8cm] {\small $g_{\mathrm{out}}$};
    \node (sisoK) [box, right of=bigamp, yshift=-.8cm] {\small  $g_{\mathrm{out}}$};
    \node (crc) [box, right of=sisoK, yshift=.8cm] {\small Check};
    \node (points) [right of=bigamp, yshift=.1cm] {$\vdots$};
    \node (list) [box, below of=crc, yshift=1cm, align=center] {\small Encoding};

    \node (sic1) [box, yshift=1cm, below of=bigamp] { \small$\mathbf{A}\hat{\mathbf{C}}_{\mathcal{S}^{(i)}}\hat{\mathbf{H}}_{\mathcal{S}^{(i)}}^{\Tr}$};
    \node (Y)[left of=sic, xshift=1cm] {\small $\mathbf{Y}^{(i-1)}$};

    \draw [arrow] (Y.east) -- (sic.west);
    \draw [arrow] (sic1.west) -| (sic.south);

    \draw [arrow] (sic.east) node[anchor=east, yshift=.3cm, xshift=1.2cm] {$\mathbf{Y}^{(i)}$}-- (bigamp);
    \draw [arrow] (bigamp.south) -| (sic1.north);

    \path [line] (bigamp.east) -- +(.5,0) |- node[anchor=center, above] {\small$p(\mathbf{c}_1\vert\mathbf{Y}^{(i)})$}(siso1.west);
    \path [line] (bigamp.east) -- +(.5,0) |- node[anchor=center, below] {\small$p(\mathbf{c}_{K^{(i)}}\vert\mathbf{Y}^{(i)})$}(sisoK.west);
    \path [line] (siso1.east) -- +(.5,0) node[anchor=center, above] {\small$\hat{W}_1$}|-  (crc.west);
    \path [line] (sisoK.east) -- +(.5,0) node[anchor=center, below] {\small$\hat{W}_{K^{(i)}}$} |-  (crc.west);

    \draw [arrow] (crc.south) -- node[anchor=center, left] {\small$\mathcal{S}^{(i)}$} (list.north);
    
    \node (S) [right of=crc, xshift=-1.4cm] {};
    \draw [arrow] (crc.east) -- node[anchor=west, yshift=.4cm, xshift=-.5cm] {\small$\hat{\mathcal{L}}^{(i+1)}=\hat{\mathcal{L}}^{(i)}\cup \mathcal{S}^{(i)}$} (S);

    \draw [arrow] (list.west) -- node[anchor=center, yshift=.3cm, xshift=0cm] {$\hat{\mathbf{C}}_{\mathcal{S}^{(i)}}$} (sic1.east);
\end{tikzpicture}
    \caption{This block diagram offers a synopsis of the decoding process.\label{fig:scheme}}
\end{figure*}
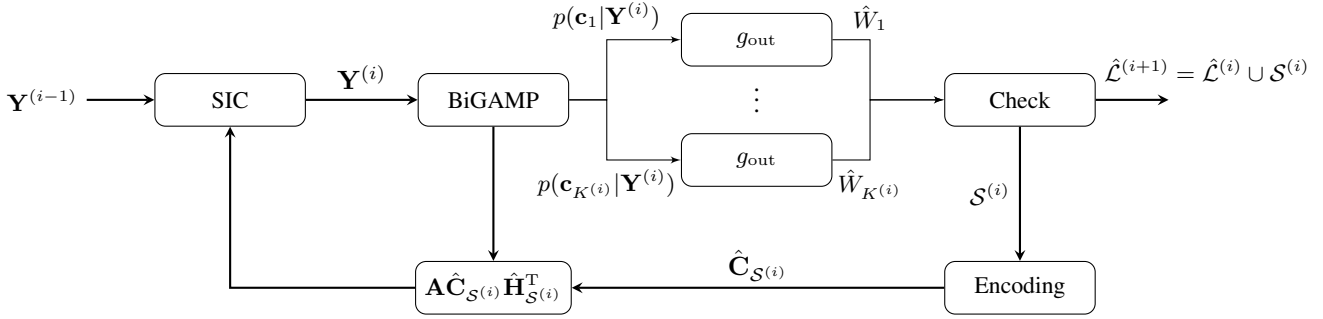

\subsection{Decoder}

In the following, we construct an iterative receiver that aims to recover the messages transmitted by the active devices. Its main components are a message-passing based detection which we drive using the \ac{BiGAMP} paradigm, a \ac{SISO} decoding for the outer channel code, and a successive interference cancellation step. In the first step the binary \ac{SPARC}-vectors $\mathbf{c}_k$ are recovered by exploiting the bilinear dependence between the \ac{MIMO} channels, \ac{SPARC}-support vectors, and receive signal. To this end, we formulate a approximate messsage passing based detection algorithm which jointly estimates the posterior distirbutions of the \ac{SPARC}-support vectors, and the \ac{MIMO} channels. The soft information of each \ac{SPARC}-support vectors is then passed to a separate instance of a \ac{SISO} decoder of the outer channel code. Decoded valid messages denoted by the set $\mathcal{S}$ are added to the list of decoded messages $\hat{\mathcal{L}}$. The detected messages a re-encoded for MMSE based channel estimation. Finally, the contribution of the decoded messages are subtracted from the receive signal, and the procedure starts a new detection round on the residual signal, iterating until the decoder cannot output a new message. The main building blocks of the algorithm are highlighted in Figure~\ref{fig:scheme}. We elaborate on the individual steps below.
\smallskip

\noindent\textit{1) BiGAMP Detector:} Consider the received signal $\mathbf{Y}$ defined in \eqref{eq:system_model}, the goal of the receiver is to detect the binary vectors $\hat{\mathrm{C}}=[\hat{\mathbf{c}}_{1},\ldots,\hat{\mathbf{c}}_{K}]$ and subsequently decode the transmitted messages to output the message list $\hat{\mathcal{L}}:=\{\hat{W}_1,\ldots,\hat{W}_{K}\}$. Provided that the coding scheme described in Section~\ref{sec:encoding}, employs an outer channel code which allows for \ac{SISO} decoding, we are interested in a probabilistic estimation of $p(\mathbf{c}_k\vert \mathbf{Y})\triangleq p(W_k\vert \mathbf{Y})$. To this end, we pose the detection problem as a Bayesian inference problem, which given \eqref{eq:system_model} is formulated over the posterior distribution of the unknown \acp{RV} $\mathbf{H}$ and $\mathbf{C}$ as
\begin{equation}
    p(\mathbf{C},\mathbf{H}|\mathbf{Y})\propto p(\mathbf{Y}|\mathbf{Z})p(\mathbf{C})p(\mathbf{H}), \label{eq:posterior_system_model}
\end{equation}
with $\mathbf{Z}=\mathbf{A}\mathbf{C}\mathbf{H}^\mathrm{T}$. The posterior distribution in \eqref{eq:posterior_system_model} is composed of the likelihood distribution of the receive signal
\begin{equation}
p(\mathbf{Y}|\mathbf{Z})= \prod\limits_{t=1}^T\prod\limits_{m=1}^M p(y_{tm}\vert z_{tm})
\end{equation}
where under the assumption of AWGN receive noise and independent antenna elements we have
\begin{equation}
p(y_{tm}\vert z_{tm})=\mathcal{CN}(z_{mt}; y_{mt},\sigma^2).
\end{equation}
For Rayleigh block fading of coherence length $T$ and independent antenna elements, the prior distribution of the unknown channels $\mathbf{H}$ is given by
\begin{equation}
p(\mathbf{H})=\prod\limits_{k=1}^{K}\prod\limits_{m=1}^Mp(h_{km})
\end{equation}
with $p(h_{km})=\CN{h_{km}}{0}{1}$.
Given the section wise partitioned code structure of \ac{SPARC} encoding, the prior distribution of the SPARC encoded binary vectors is given by the \ac{pmf}
\begin{equation}
p(\mathbf{C})=\prod\limits_{k=1}^{K}\prod\limits_{l=1}^{L} p(\mathbf{c}_k^{(l)})
\end{equation}
where given the section-wise encoding structure of \acp{SPARC} \eqref{eq:sparc_coder} we have
\begin{equation}
p(\mathbf{c}_k^{(l)})=\frac{1}{Q}\sum\limits_{n=1}^{Q}\delta(c_{nk}^{(l)}-1)\prod\limits_{j\neq n} \delta(c_{jk}^{(l)}).\label{eq:subvector_pmf}
\end{equation}
Given the prior distributions, the quantity of interest is the marginal posterior distribution of the binary entries 
\begin{equation}
p(c_{nk}^{(l)}|\mathbf{Y})\propto\sum_{\mathbf{C}/c_{nk}^{(l)}}\int_{d\mathbf{H}}p(\mathbf{Y}|\mathbf{Z})p(\mathbf{C})p(\mathbf{H})\label{eq:marginal_posterior}
\end{equation}
where $\mathbf{C}/c_{nk}^{(l)}$ denotes the matrix $\mathbf{C}$ with omitted element $c_{nk}^{(l)}$. The marginalized posterior probability of the binary vector elements can be translated to the posterior distribution of code symbols as  
\begin{equation}
    p(c_{nk}^{(l)}=1|\mathbf{Y})\triangleq p(W_k^{(l)}=n|\mathbf{Y})    
\end{equation}
which we passed in a subsequent iteration step as symbol priors to a soft-input-soft-output (SISO) decoder of the outer channel code. Unfortunately, the considered system model renders the computation of \eqref{eq:marginal_posterior} intractable. To reduce complexity, we resort to approximate message passing based scheme to solve this problem. To be specific, in the following we will resort to the bilinear generalized approximate message passing algorithm (BiGAMP) \cite{parker_bilinear_2014} which we extend to fit our system model. We provide the derivation of the message updates in Appendix~\ref{sec:message-derivation} and summarize the overall detection algorithm in Algorithm~\ref{alg:bigamp}. 

\smallskip

\noindent\textit{2) SISO Decoding:} The decoding step performs parallel \ac{SISO} decoding over the posterior distributions of the detected \ac{SPARC} vectors, i.e., $\{p(\mathbf{c}_k\vert \mathbf{Y}^{(i)})\}_{k\in[K^{(i)}]}$. After completion, messages are decoded and checked for validity. Valid messages denoted by the set $\mathcal{S}^{(i)}$ are added to the list of decoded messages $\hat{\mathcal{L}}^{(i+1)}=\hat{\mathcal{L}}^{(i)}\bigcup \mathcal{S}^{(i)}$. Furthermore, the detected messages are $\mathcal{S}^{(i)}$ re-encoded and passed with the respective channel estimations to the \ac{SIC} step.

\smallskip
\noindent\textit{3) SIC:} In the final step of the proposed iterative detection and decoding scheme, successfully decoded messages are leveraged to mitigate interference through \ac{SIC}. This is accomplished by utilizing the respective estimated channels $\hat{\mathbf{H}}_{\mathcal{S}}$ for all the users for which the previous decoding step has produced valid messages. The re-encoded valid messages together with the respective channel estimates and the shared dictionary are utilized to subtract their contribution from the received signal, i.e.,
\begin{equation}
    \mathbf{Y}^{(i+1)}=\mathbf{Y}^{(i)}-\mathbf{A}\hat{\mathbf{C}}_{\mathcal{S}^{(i)}}\hat{\mathbf{H}}_{\mathcal{S}^{(i)}}^{\Tr}.\label{eq:residual}
\end{equation}
The residual signal \eqref{eq:residual} is then passed to the detection step for another round of detection and decoding. The whole iterative process is repeated until no valid messages are produced by the decoding step. We note, that by subtracting the contribution of the detected messages from the receive signal, we implicitly reduced the size of the detection problem from $K$ to $K^{(i+1)}=K^{(i)}-\vert\mathcal{S}^{(i)}\vert$. As a consequence, the computational complexity is reduced after each successful detection round.

\subsection{Complexity}
The computational complexity of the iterative detection scheme described in Figure~\ref{fig:scheme} is mainly influenced by the complexity of the \ac{BiGAMP} based detection step and the parallel \ac{SISO} decoding. Assuming sequences from a structured dictionary are used, e.g., Gabor frames \cite{agostini_constant_2022} or sub-sampled FFT matrices \cite{fengler_pilot-based_2022}, the complexity of the detection step is in the order of $\mathcal{O}(KNMT\log T)$. The complexity of the outer channel \ac{SISO} decoding step depend on the employed code and decoder. Its contribution is however independent of $K$, since decoding can be performed in parallel.

\begin{algorithm}[h]
\caption{BiGAMP for Detection. }\label{alg:bigamp}
\allowdisplaybreaks
\begin{algorithmic}[1]
\renewcommand{\algorithmicrequire}{\textbf{Input:}}
\renewcommand{\algorithmicensure}{\textbf{Output:}}
\STATE \textbf{Input}: $\mathbf{Y}$, $\mathbf{A}$, $t_{\mathrm{Max}}$
\STATE \textbf{Initialize:}\\
$\forall (n,k):\forall(n,k):\hat{c}_{nk}\sim\mathrm{Bernoulli}(\frac{1}{Q})$\\
$\forall (k,m):\hat{h}_{nk}(0)=0,\;\forall (k,m):\hat{v}_{nk}^h(0)=0$\\
\FOR {$i=1$ to $T_{\mathrm{Max}}$} 
\STATE $\theta_{nm}^{(i)}=\sum_{t=1}^T a_{tn}^*\mu_{tm}^{z,(i)}+\mu_{nm}^{x,(i)}$\\
\STATE $\mu_{nm}^{x,(i)}=\frac{\theta_{nm}^{(i)}\nu_{nm}^{x,(i)}+\mu_{nm}^{x,(i)}\nu_{nm}^{z,(i)}}{\nu_{nm}^{x,(i)}+\nu_{nm}^{z,(i)}}$\\
\STATE $\nu_{nm}^{x,(i)}=\frac{\nu_{nm}^{z,(i)}\nu_{nm}^{x,(i)}}{\nu_{nm}^{z,(i)}+\nu_{nm}^{x,(i)}}$\\
\STATE $\nu_{nm}^{z,(i)}=\sigma^2+\frac{1}{T}\sum_{n=1}^N\nu_{nm}^{x,(i+1)}$\\
\STATE $\mu_{tm}^{z,(i)}=y_{tm}-\sum_{n}a_{tn}\mu_{nm}^{x,(i)}$\\
\STATE $\mu_{tm}^{z,(i)}=\mu_{tm}^{z,(i)}+\frac{\mu_{tm}^{z,(i)}}{T}\sum_{n=1}^NF(\theta_{nm}^{(i)},\nu_{nm}^{z,(i)})$\\
\STATE $\nu^{p,(i)}_{nm}=\sum_k\vert\mu^{c,(i)}_{nk}\vert^2\nu^{h,(i)}_{km}+\nu^{c,(i)}_{nk}\vert \mu^{h,(i)}_{km}\vert^2\nonumber+\nu^{c,(i)}_{nk}\nu^{h,(i)}_{km}$\\
\STATE $\mu^{p,(i)}_{nm}=\sum_k \mu^{c,(i)}_{nk}\mu^{h,(i)}_{km}-\mu^{\alpha,(i)}_{nk}\vert\mu^{c,(i)}_{nk}\vert^2 \nu^{h,(i)}_{km}$\\
\STATE $\nu_{nm}^{\alpha,(i)}=(\nu^{p,(i)}_{nm}+\nu_{nm}^{x,(i)})^{-1}$\\
\STATE $\mu^{\alpha,(i)}_{nm}=\frac{\mu_{nm}^{x,(i)}-\mu^{p,(i)}_{nm}(i)}{\nu^{p,(i)}_{nm}+\nu^{x,(i)}_{nm}}$\\
\STATE $\nu_{km}^{r,(i)}=(\sum_n \vert\mu^{c,(i)}_{nk}\vert^2\nu_{nm}^{\alpha,(i)})^{-1}$\\
\STATE $\mu^{r,(i)}_{km}=\mu^{h,(i)}_{km}(1-\nu^{r,(i)}_{km}\sum_n \nu^{c,(i)}_{nk}\nu_{nm}^{\alpha,(i)})$\\
\STATE $\mu^{r,(i)}_{km}=\mu^{r,(i)}_{km}+\nu^{r,(i)}_{km}\sum_n\mu^{c,(i),*}_{nk}\mu^{\alpha,(i)}_{nm}$\\
\STATE $\nu^{q,(i)}_{nk}=(\sum_m\vert\mu^{h,(i)}_{km}\vert^2\nu_{nm}^{x,(i)})^{-1}$\\
\STATE $\mu^{q,(i)}_{nk}=\mu^{c,(i)}_{nk}(1-\nu_{nk}^{q,(i)}\sum_m \nu^{h,(i)}_{km}\nu_{nm}^{x,(i)})$\\
\STATE $\mu^{q,(i)}_{nk}=\mu^{q,(i)}_{nk}+\nu_{nk}^{q,(i)}\sum_m\mu^{h,(i),*}_{km}\mu^{x,(i)}_{nm}$
\STATE $\mu^{h,(i+1)}_{km}=\eqref{eq:up_h_mu},\quad\nu_{km}^{h,(i+1)}=\eqref{eq:up_h_var}$\\
\STATE $\mu^{c,(i+1)}_{nk}=\eqref{eq:up_c_mu},\quad\nu_{nk}^{c,(i+1)}=\eqref{eq:up_c_var}$
\ENDFOR
\end{algorithmic}  
\end{algorithm}

\section{Numerical Results}

We compare BiSPARCs with the state-of-the-art schemes proposed by Fengler et al. \cite{fengler_pilot-based_2022}, Gkagkos et al. \cite{gkagkos_fasura_2022}, and Decurninge et al. \cite{decurninge_tensor-based_2020}, to assess its performance. To ensure a fair comparison between the two \ac{URA} communication schemes, we pick parameters for our system that match their reported implementation. Specifically, we target a system with $B\approx100$ message bits, $T=3200$ complex channel uses, and a target probability of error $P_e\leq0.05$. 
\smallskip

\noindent\textit{1) BiSPARCs:} For the concatenated code we choose a 5G NR compliant binary LDPC code \cite{3gpp_nr_2017} parametrized by $(n=110,k=100)$, leading to a SPARC parametrization of $(L=14,Q=256)$. The total amount of transmitted information amounts to $B=100$-bit. For the shared dictionary we choose a Gabor frame construction based on an Alltop seed vector \cite{agostini_constant_2022}, which exhibits excellent coherence properties and allows multiplication to be realized efficiently as FFT operations.
\smallskip

\noindent\textit{2) Pilot-Based (Fengler et al. \cite{fengler_pilot-based_2022}):} The pilot-based approach is parametrized with pilots drawn from a randomly sub-sampled Fourier matrix and a polar code with a \ac{SCL} decoder with $16$ \ac{CRC} bits and a list size of $32$.
\smallskip

\noindent\textit{3) FASURA (Gkagkos et al. \cite{gkagkos_fasura_2022}):} The parameters for FASURA are $n_p=896$, $L=9$, $n_c=512$, $nL = 64$ and $J = 216$. Spreading sequence matrices and pilot matrix are randomly selected. Elements of the spreading sequence matrices are drawn from an i.i.d. complex Gaussian distribution with zero mean and variance $1$, and normalized to unit power. For $K = 100$, $12$ CRC bits, whereas when $K\geq 100$, we employ $16$ CRC bits, are used. 
\smallskip

\noindent\textit{4) TBM (Decurninge et al. \cite{decurninge_tensor-based_2020}):} The TBM approach is parameterized with a tensor signature $(8,5,5,4,4)$ and an outer BCH code. The values have been obtained with the higher value $P_e=0.1$.

\begin{figure}[h!]
	\centering
	\includegraphics[width=1\columnwidth]{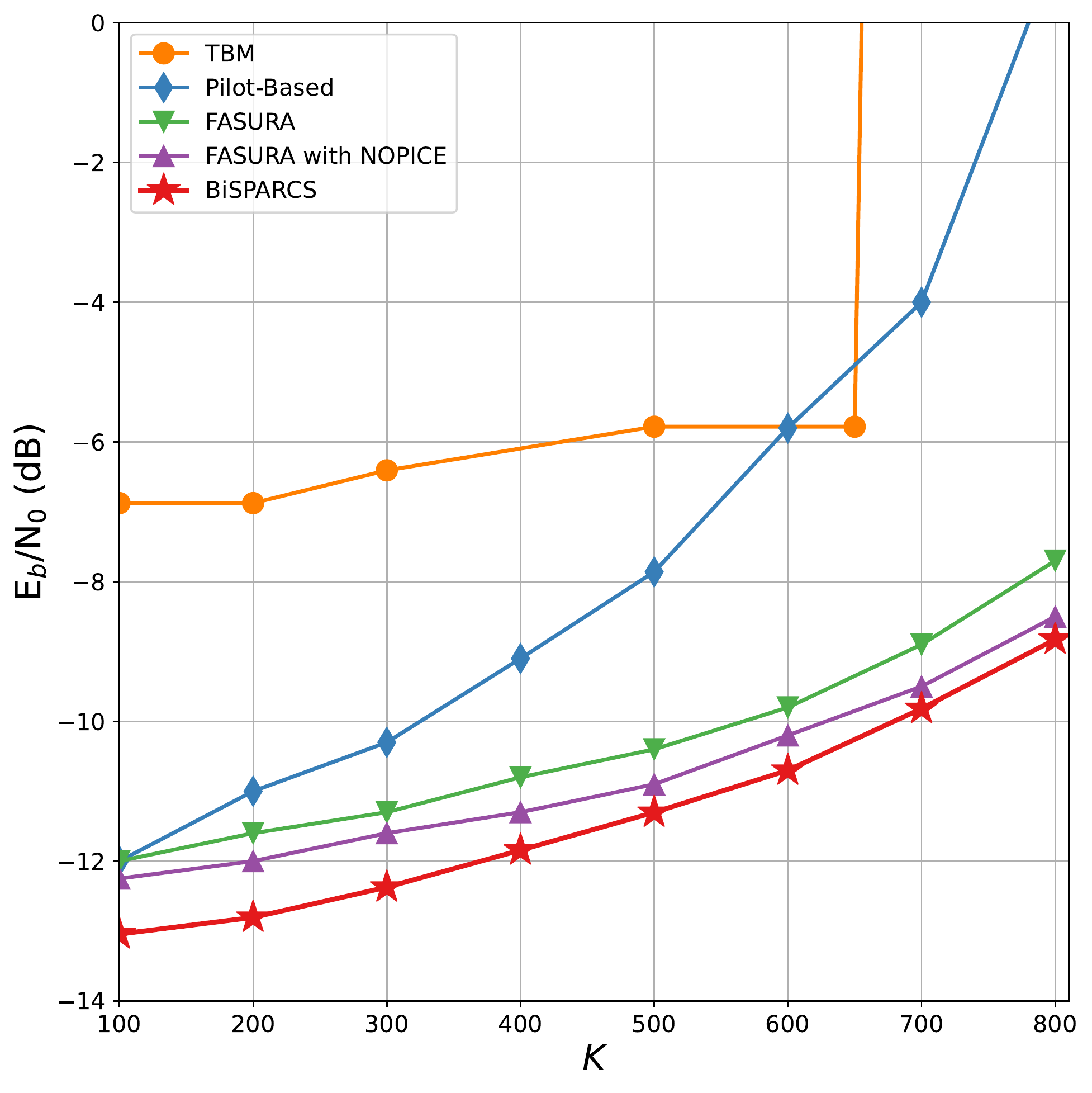}
\caption{Required $E_b/N_{0}$ for achieving target error probabilities $P_e$ at fixed $T=3200$. Proposed approach is compared against TBM \cite{decurninge_tensor-based_2020}, Pilot-Based \cite{fengler_massive_2019}, FASURA \cite{gkagkos_fasura_2022}, and FASURA with NOPICE \cite{gkagkos_fasura_2022}.}\label{fig:unsourced_100_bits_50_M}
\end{figure}

Fig.~\ref{fig:unsourced_100_bits_50_M} shows the required $E_b/N_0=PT/(B\sigma^2)$, to achieve the target $P_e$ for the number of antennas at the base station of $M=50$. We plot the performance of BiSPARCs, along with that of the other competing schemes. For the operational parameters studied and a user population exceeding $K\geq100$ active devices, BiSPARCs outperforms all state-of-the-art schemes. Interestingly, as the number of active users grows, the gap between BiSPARCs and the best performing competing scheme, i.e., FASURA with NOPICE, decreases, showing similar scalability towards increasing number of active devices starting from $K>400$. We furthermore remark that the performance of the FASURA with NOPICE is achieved via an additional MMSE step to increase the accuracy of the channel estimation, which significantly increases its complexity. However, BiSPARCs on the other hand provides an implicit MMSE channel estimate as part of the detection, which mitigates the need for additional channel estimation keeping the overall complexity under control. 

\begin{figure}[h!]
\centering
\includegraphics[width=1\columnwidth]{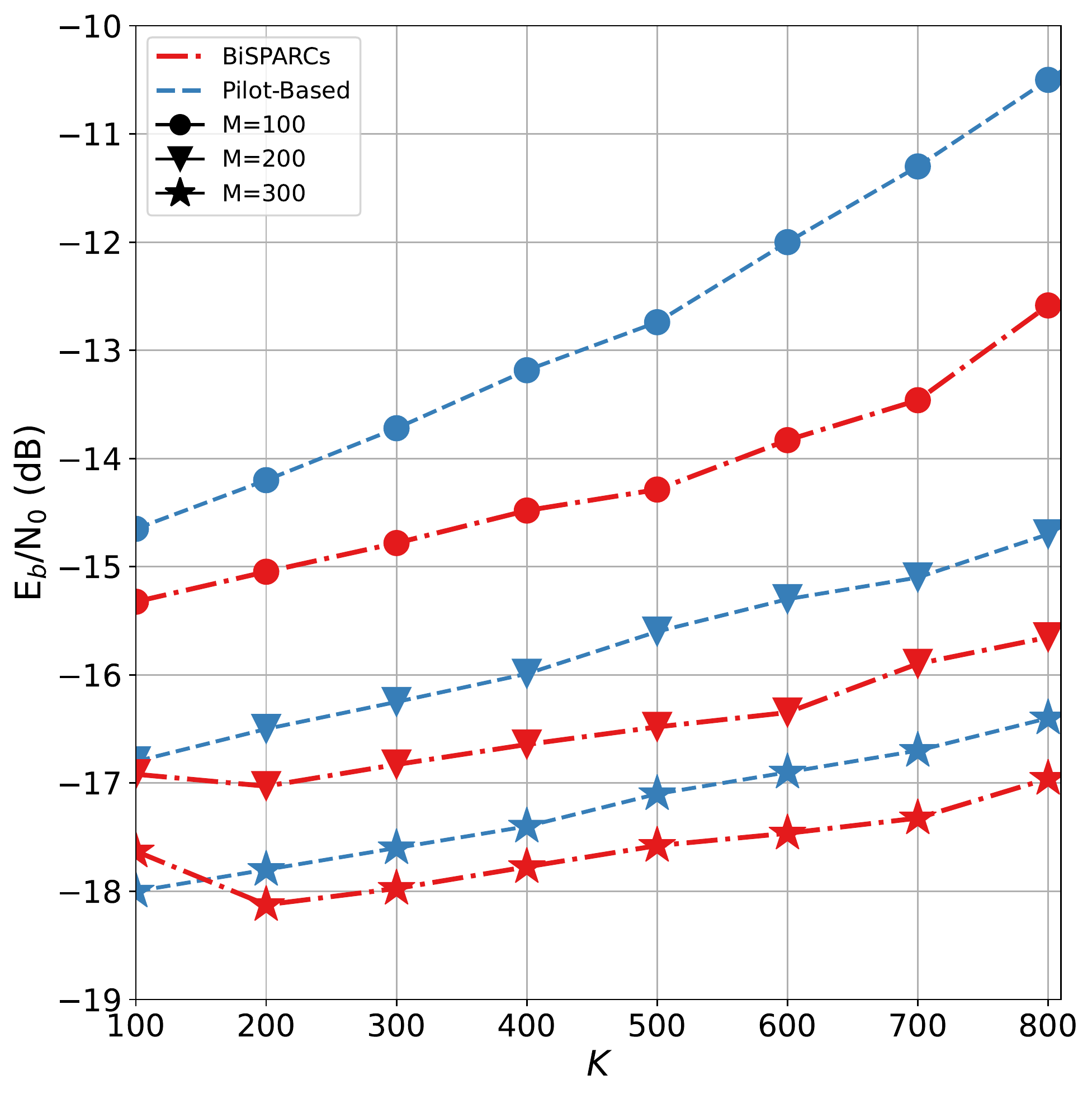}
\caption{Required $E_b/N_{0}$ for achieving target error probability $P_e\leq0.05$ at fixed $T=3200$ for different antenna sizes $M\in\{100,200,300\}$. We compare BiSPARCs against the values reported for the pilot-based approach in \cite{fengler_massive_2019}.}
\label{fig:unsourced_100_bits}
\end{figure}

In Fig.~\ref{fig:unsourced_100_bits} we compare the performance of BiSPARCs against the values reported for the pilot-based approach in \cite{fengler_pilot-based_2022} for antenna sizes $M>50$, i.e., $M\in\{100,200,300\}$. Compared to the baseline, we observe an overall superior performance and better scaling property towards increasing number of active users. 

\section{Conclusion}
In this paper, we presented a coding scheme for \ac{URA} in the quasi-static Rayleigh fading with a massive MIMO receiver. We proposed a novel communication scheme called BiSPARCs which has competitive performance with existing schemes for $K\geq 100$. BiSPARCs leverages the bilinear relation between the MIMO channels and a \ac{SPARC} based \ac{URA} coding scheme which arises in the considered massive MIMO regime. This relation allows for joint \ac{SPARC} decoding and channel estimation, for which we formulated a \ac{BiGAMP} based message passing receiver. By concatenating the detected \ac{SPARC} vectors with an outer LDPC based channel code, the messages for every device can be decoded in parallel, which mitigates the need for additional coding for user separation.
 
\section*{Acknowledgment}
The authors acknowledge the financial support by the Federal Ministry of Education and Research of Germany in the program of “Souverän. Digital. Vernetzt.” joint project 6G-RIC (BMBF-16KISK020K). The authors would like to acknowledge the contributions of their colleagues in the projects, although the views expressed in this contribution are those of the authors and do not necessarily represent the projects.
\appendices
\section{Derivation of Message Approximations}\label{sec:message-derivation}

In the following, we derive the message passing updates for the Bayesian joint detection of the channel matrix $\mathbf{H}$ and \ac{SPARC} support matrix $\mathbf{C}$. To this end we use $\mathbf{c}_n=\mathbf{C}_n^{\mathrm{T}}$ to denote the $n$-th row of $\mathbf{C}$, $\mathbf{h}_m=\mathbf{H}_m^{\mathrm{T}}$ to denote the $m$-th row of $\mathbf{H}$, and $\mathbf{a}_t=\mathbf{A}^{\mathrm{T}}_t$ to denote the $t$-th row of the dictionary $\mathbf{A}$. We furthermore introduce the auxiliary latent variable $\mathbf{X}\in\mathbb{C}^{N\times M}$. To derive the message passing updates, we proceed by stating the posterior distribution of the unknown \acp{RV}, i.e., $\mathbf{H}$, $\mathbf{C}$ and $\mathbf{X}$, given then receive signal $\mathbf{Y}$ which reads as
\begin{multline}
    p(\mathbf{H},\mathbf{C},\mathbf{X}\vert \mathbf{Y})\propto \\
    \prod_{t,m,n} p(y_{tm}\vert \mathbf{A}\mathbf{x}_m)\delta(x_{nm}-\mathbf{c}_n^{\mathrm{T}}\mathbf{h}_m)p(\mathbf{c}_n)p(\mathbf{h}_m).\label{eq:posterior_app}
\end{multline}
\ac{BP} is an iterative procedure to find the fixed point solution of the following closed equations for the distributions defined on the edges of the factor graph representation of \eqref{eq:posterior_app}. The factorization of \eqref{eq:posterior_app} allows for a separate consideration of the \acp{RV} $\mathbf{X}$, $\mathbf{C}$, and $\mathbf{H}$ in the formulation of the message updates. We start by formulating the message updates for the auxiliary latent variable $\mathbf{X}$ which relate to the affine mapping part of the system model, followed by the derivation of the message updates for the \acp{RV} involved in the bilinear part of the system model, i.e., $\mathbf{H}$ and $\mathbf{C}$. 
\subsection{Affine Message Updates}
The \ac{BP} message updates for the affine part of the system model are given by
\begin{align}
\M{tm}{nm}{(i+1)}{x_{nm}}&\propto\int\limits_{d\mathbf{x}_{m/n}}p(y_{tm}\vert\mathbf{a}_t^{\mathrm{T}}\mathbf{x}_m)\prod_{q\neq n}\M{qm}{tm}{(i)}{x_{qm}}\label{eq:message_tm_nm}\\
\M{nm}{tm}{(i+1)}{x_{nm}}&\propto\M{nm}{nm}{(i)}{x_{nm}}\prod_{v\neq t}\M{tm}{nm}{(i)}{x_{nm}}\label{eq:message_nm_tm}
\end{align}
To render to calculations of the message updates tractable, we proceed by approximating \eqref{eq:message_nm_tm} with a Gaussian - which can be justified in the large system limit (i.e., $N,\;T\rightarrow\infty$ with $T/N$ fixed) using the \ac{CLT} argument. The \textit{Gaussianity} of the likelihood function in \eqref{eq:message_tm_nm} implies that the respective message update will also be Gaussian. In particular, the message approximation for \eqref{eq:message_nm_tm} can be stated as 
\begin{equation}
    \M{nm}{tm}{(i+1)}{x_{nm}}\approx\mathcal{CN}(x_{nm};\mu_{nm,t}^{x,(i)},\nu_{nm,t}^{x,(i)}).\label{eq:approx_nm_tm}
\end{equation}
Using the known fact about Gaussian products, i.e.,
\begin{equation}
    \prod_q\mathcal{CN}(\theta;\mu_q,\nu_q)\propto\mathcal{CN}\left(\theta;\frac{\sum_q\mu_q/\nu_q}{\sum_q1/\sigma_q},\frac{1}{\sum_q1/\nu_q}\right).\label{eq:gaussian_fact1}
\end{equation}
and \eqref{eq:approx_nm_tm} leads to following approximations
\begin{subequations}
\begin{align}
\M{tm}{nm}{(i)}{x_{nm}}&\approx \mathcal{CN}\left(a_{tn}x_{nm};\mu_{tm,n}^{z,(i)},\nu_{tm,n}^{z,(i)}\right)\\
\mu_{tm,n}^{z,(i)}&\triangleq y_{tm}-\sum_{q\neq n}a_{tq}\mu_{qm,t}^{x,(i)}\\
\nu_{tm,n}^{z,(i)}&\triangleq \sigma^2+\sum_{q\neq n}\vert a_{tq}\vert^2\nu_{qm,t}^{x,(i)},
\end{align}
\end{subequations}
and
\begin{align}
    \M{nm}{nm}{(i+1)}{x_{nm}}&\approx\mathcal{CN}(x_{nm};\mu_{nm}^{x,(i)},\nu_{nm}^{x,(i)})\label{eq:approx_nm_nm1}.
\end{align}
From \eqref{eq:approx_nm_tm}, we see that $\mu_{nm,t}^{x,(i+1)}$ and $\nu_{nm,t}^{x,(i+1)}$ are determined by the mean and variance, respectively, of the pdf 
\begin{multline}
    \M{nm}{tm}{(i+1)}{x_{nm}}\propto\\
    \M{nm}{nm}{(i)}{x_{nm}}\prod_{v\neq t}\M{tm}{nm}{(i)}{x_{nm}}\label{eq:approx_nm_tm1}
\end{multline}
where the product term can be reformulated using \eqref{eq:gaussian_fact1} to 
\begin{subequations}
\begin{align}
\prod_{l\neq t}\M{lm}{nm}{(i)}{x_{nm}}&\propto\mathcal{CN}(x_n;\mu_{tm,n}^{z,(i)},\nu_{tm,n}^{z,(i)})\\
\mu_{tm,n}^{z,(i)}&=\sum_{l\neq t}a_{ln}^{*}\mu_{lm,n}^{z,(i)}\\
\nu_{tm,n}^{z,(i)}&=\sigma^2+\frac{1}{T}\sum_{t=1}^T\nu_{nm,t}^{x,(i)}.
\end{align}    \label{eq:msg_lm_nm_prod}
\end{subequations}
Using a first-order \ac{AMP}-style approximation, we can greatly simplify the update steps involved in \eqref{eq:msg_lm_nm_prod}, by tracking only $\mathcal{O}(N)$ variables.
\begin{subequations}
    \begin{align}
        \theta_{nm}^{(i)}={}&\sum_{t=1}^T a_{tn}^*\mu_{tm}^{z,(i)}+\mu_{nm}^{x,(i)}\\
        \mu_{nm}^{x,(i+1)}={}&\frac{\theta_{nm}^{(i)}\nu_{nm}^{x,(i)}+\mu_{nm}^{x,(i)}\nu_{nm}^{z,(i)}}{\nu_{nm}^{x,(i)}+\nu_{nm}^{z,(i)}}\\
        \nu_{nm}^{x,(i+1)}={}&\frac{\nu_{nm}^{z,(i)}\nu_{nm}^{x,(i)}}{\nu_{nm}^{z,(i)}+\nu_{nm}^{x,(i)}}\\
        \nu_{nm}^{z,(i+1)}={}&\sigma^2+\frac{1}{T}\sum_{n=1}^N\nu_{nm}^{x,(i+1)}\\
        \mu_{tm}^{z,(i+1)}={}&y_{tm}-\sum_{n}a_{tn}\mu_{nm}^{x,(i)}\nonumber\\
        &+\frac{\mu_{tm}^{z,(i)}}{T}\sum_{n=1}^NF(\theta_{nm}^{(i)},\nu_{nm}^{z,(i)})
    \end{align}\label{eq:amp_pr}
\end{subequations}
with
\begin{multline}
     F(\mu_{tm}^{x,(i)},\nu_{tm}^{x,(i)})=\frac{\nu_{nm}^{x,(i)}}{\nu_{tm}^{x,(i)}+\nu_{nm}^{x,(i)}}\\
     +\frac{(\nu_{nm}^{x,(i)})^2\vert\mu_{tm}^{x,(i)}\vert^2}{\nu_{tm}^{x,(i)}(\nu_{tm}^{x,(i)}+\nu_{nm}^{x,(i)})^2}
\end{multline}
   
\subsection{Bilinear Message Updates}
With \eqref{eq:approx_nm_nm1} and the message updates from \eqref{eq:amp_pr} we proceed by deriving the message updates for $c_{nk}$ and $h_{mk}$. To this end we first pass the message updates from the \ac{RV} $x_{nm}$ to $\mathbf{c}_n^{\Tr}\mathbf{h}_m$, leading to the likelihood calculation
\begin{multline} 
\M{nm}{nm}{(i)}{\mathbf{c}_n,\mathbf{h}_m}\propto\\
\int\limits_{dx_{nm}}\M{nm}{nm}{(i)}{x_{nm}}\delta(x_{nm}-\mathbf{c}_n^{\mathrm{T}}\mathbf{h}_m)
\end{multline}
which using \eqref{eq:approx_nm_nm1} leads to
\begin{equation}
    \M{nm}{nm}{(i+1)}{\mathbf{c}_n,\mathbf{h}_m}\propto\mathcal{CN}(\mathbf{c}_n^{\mathrm{T}}\mathbf{h}_m;\mu_{nm}^{x,(i)},\nu_{nm}^{x,(i)}).
\end{equation}
Following the BiGAMP derivation procedure in \cite{parker_bilinear_2014}, we introduce the auxiliary \ac{RV}, i.e., $p_{nm}=\mathbf{c}_n^{\Tr}\mathbf{h}_m$, for which we the message updates are given by
\begin{subequations}
\begin{align}
\nu^{p,(i)}_{nm}{}=&\sum_k\vert\mu^{c,(i)}_{nk}\vert^2\nu^{h,(i)}_{km}+\nu^{c,(i)}_{nk}\vert \mu^{h,(i)}_{km}\vert^2\nonumber\\
&+\nu^{c,(i)}_{nk}\nu^{h,(i)}_{km}\\
\mu^{p,(i)}_{nm}{}=&\sum_k \mu^{c,(i)}_{nk}\mu^{h,(i)}_{km}-\mu^{\alpha,(i)}_{nk}\vert\mu^{c,(i)}_{nk}\vert^2 \nu^{h,(i)}_{km}.
\end{align}
\end{subequations}
To ease further derivation we follow by introducing the helper variable $\alpha_{nm}$ with moment updates given by
\begin{subequations}
\begin{align}
\nu_{nm}^{\alpha,(i)}&=(\nu^{p,(i)}_{nm}+\nu_{nm}^{x,(i)})^{-1},\\
\mu^{\alpha,(i)}_{nm}&=\frac{\mu_{nm}^{x,(i)}-\mu^{p,(i)}_{nm}(i)}{\nu^{p,(i)}_{nm}+\nu^{x,(i)}_{nm}}.
\end{align}\label{eq:alpha}
\end{subequations}
Using the moment updates from \eqref{eq:alpha}, the moments of the likelihood for $h_{km}$ and $c_{nk}$ are are given by
\begin{subequations}
\begin{align}
\Delta_{km}^{(i)}(h_{km})&=\mathcal{CN}(h_{km};\mu^{r,(i)}_{km},\nu_{km}^{r,(i)})\\
 \nu_{km}^{r,(i)}&=(\sum_n \vert\mu^{c,(i)}_{nk}\vert^2\nu_{nm}^{\alpha,(i)})^{-1}\\
 \mu^{r,(i)}_{km}{}&=\mu^{h,(i)}_{km}(1-\nu^{r,(i)}_{km}\sum_n \nu^{c,(i)}_{nk}\nu_{nm}^{\alpha,(i)})\nonumber\\
 &+\nu^{r,(i)}_{km}\sum_n\mu^{c,(i),*}_{nk}\mu^{\alpha,(i)}_{nm}.
\end{align}\label{eq:D_h_km}
\end{subequations}
and
\begin{subequations}
\begin{align}
\Delta_{nk}^{(i)}(c_{nk})={}&\mathcal{CN}(c_{nk};\mu^{q,(i)}_{nk},\nu_{nk}^{q,(i)})\\
\nu^{q,(i)}_{nk}={}&(\sum_m\vert\mu^{h,(i)}_{km}\vert^2\nu_{nm}^{x,(i)})^{-1},\\
\mu^{q,(i)}_{nk}={}&\mu^{c,(i)}_{nk}(1-\nu_{nk}^{q,(i)}\sum_m \nu^{h,(i)}_{km}\nu_{nm}^{x,(i)})\nonumber\\
&+\nu_{nk}^{q,(i)}\sum_m\mu^{h,(i),*}_{km}\mu^{x,(i)}_{nm}.
\end{align}\label{eq:D_c_nk}
\end{subequations}
With \eqref{eq:D_h_km} and \eqref{eq:D_c_nk}, the final \ac{MMSE} estimates for $h_{mk}$ and $c_{nk}$ read as
\begin{subequations}
    \begin{align}
        \mu^{h,(i+1)}_{km}&=\int\limits_{h_{km}}h_{km}\Delta_{h_{km}}^{(i)}(h_{km})p(h_{km})\\
        \nu_{km}^{h,(i+1)}&=\int\limits_{h_{km}}\vert h_{km}\vert^2\Delta_{h_{km}}^{(i)}(h_{km})p(h_{km})-\vert\mu^{h,(i+1)}_{km}\vert^2
    \end{align}\label{eq:mmse_hkm}
\end{subequations}
and
\begin{subequations}
    \begin{align}
        \mu^{c,(i+1)}_{nk}&=\int\limits_{c_{nk}}c_{nk}\Delta_{c_{nk}}^{(i)}(c_{nk})p(c_{nk})\\
        \nu_{nk}^{c,(i+1)}&=\int\limits_{c_{nk}}\vert c_{nk}\vert^2\Delta_{c_{nk}}^{(i)}(c_{nk})p(c_{nk})-\vert\mu^{c,(i+1)}_{nk}\vert^2.
    \end{align}\label{eq:mmse_cnk}
\end{subequations}
We provide the derivations of \eqref{eq:mmse_hkm} and \eqref{eq:mmse_cnk} with respect to the prior distributions of our system model in Appendix~\eqref{sec:mmse_updates}.

\section{MMSE Updates}\label{sec:mmse_updates}
In the following, we derive the \ac{MMSE} updates in \eqref{eq:mmse_hkm} and \eqref{eq:mmse_cnk} assuming the prior distributions from Section~\ref{sec:encoding}. 
\smallskip

\noindent\textit{1) Channel Estimates:} Given the prior of the i.i.d. Rayleigh fading, i.e., $h_{km}\propto\mathcal{CN}(h_{km};0,1)$, the Gaussian likelihood $\Delta_{h_{km}}^{(i)}(h_{km})=\mathcal{CN}(h_{km};\mu^{r,(i)}_{km},\nu_{km}^{r,(i)})$ and fact \eqref{eq:gaussian_fact1}, we get
\begin{multline}
    \mathcal{CN}(h_{km};\mu^{r,(i)}_{km},\nu_{km}^{r,(i)})\mathcal{CN}(h_{km};0,1)\propto\\
    \mathcal{CN}\left(h_{km};\frac{\mu^{r,(i)}_{km}}{\nu_{km}^{r,(i)}+1},\frac{\nu_{km}^{r,(i)}}{\nu_{km}^{r,(i)}+1}\right)
\end{multline}
from where it is not hard to see the solution which reads as
\begin{subequations}
    \begin{align}
        \mu^{h,(i+1)}_{km}&=\frac{\mu^{r,(i)}_{km}}{\nu_{km}^{r,(i)}+1}\label{eq:up_h_mu}\\
        \nu_{km}^{h,(i+1)}&=\frac{\nu_{km}^{r,(i)}}{\nu_{km}^{r,(i)}+1}\label{eq:up_h_var}.
    \end{align}
\end{subequations}

\noindent\textit{2) SPARC Estimates:} To derive \eqref{eq:mmse_cnk}, we use the section-wise prior distribution of the \ac{SPARC} vectors, which we restate here
 \begin{equation}
     p(\mathbf{c}_k^{(l)})=\frac{1}{Q}\sum\limits_{n=1}^{Q}\delta(c_{nk}^{(l)}-1)\prod\limits_{j\neq n} \delta(c_{jk}^{(l)}).\label{eq:subvector_pmf_1}
 \end{equation}
 We proceed by marginalizing \eqref{eq:mmse_cnk} with respect to $\mathbf{c}_k^{(l)}/c_{nk}^{(l)}$, i.e.,
 \begin{multline}
    p(c_{nk}^{(l)}\vert \boldsymbol{\mu}_{nk}^{q,(l)},\boldsymbol{\nu}_{nk}^{q,(l)})=\\
    \sum\limits_{\mathbf{c}_k^{(l)}/c_{nk}^{(l)}}\frac{1}{Q}\sum\limits_{n=1}^{Q}\delta(c_{nk}^{(l)}-1)\prod\limits_{j\neq n} \delta(c_{jk}^{(l)})\prod\limits_{n} \Delta_{c_{nk}^{(l)}}^{(i)}(c_{nk}^{(l)})
 \end{multline} 
 which after some algebra is given by
 \begin{multline}
    p(c_{nk}^{(l)}\vert \boldsymbol{\mu}_{nk}^{q,(l)},\boldsymbol{\nu}_{nk}^{q,(l)})=\\
    \begin{cases}
        \frac{1}{Q}\Delta_{c_{nk}^{(l)}}^{(i)}(1)\prod\limits_{j\neq n} \Delta_{c_{jk}^{(l)}}^{(i)}(0)&\mathrm{for}\quad c_{nk}^{(l)}=1\\
        \frac{1}{Q}\sum\limits_{q\neq n}\Delta_{c_{qk}^{(l)}}^{(i)}(1)\prod\limits_{j\neq q} \Delta_{c_{jk}^{(l)}}^{(i)}(0)&\mathrm{for}\quad c_{nk}^{(l)}=0.\label{eq:marg_pdf}
    \end{cases}
 \end{multline}
Given \eqref{eq:marg_pdf}, the \ac{MMSE} estimates from \eqref{eq:mmse_cnk} can now be easily computed via
\begin{subequations}
\begin{align}
    \mu^{c,(i+1),(l)}_{nk}&=(1+\exp{\frac{2\mu_{nk}^{q,(l)}-1}{\nu_{nk}^{q,(l)}}}\sum_{r\neq n}\exp{\frac{1-2\mu_{rk}^{q,(l)}}{\nu_{rk}^{q,(l)}}})^{-1}\label{eq:up_c_mu}\\
    \nu^{c,(i+1),(l)}_{nk}&=\vert\mu^{c,(i+1),(l)}_{nk}\vert^2+\nu_{nk}^{q,(l)}\label{eq:up_c_var}.
\end{align}
\end{subequations}

\bibliographystyle{IEEEtran}
\bibliography{IEEEabrv,./bibliography/unsourced-random-access}
\end{document}